\documentclass[sigconf,nonacm]{acmart}
\AtBeginDocument{%
  }

\setcopyright{none}
\copyrightyear{2026}
\acmYear{2026}
\acmDOI{}
\acmConference[ASE 2026]{Proceedings of the 41st IEEE/ACM International Conference on Automated Software Engineering}{November 2026}{Seoul, South Korea}
\acmISBN{}

\usepackage{quoting}
\usepackage{xspace}
\usepackage{subcaption}

\newcommand{\toolname}{SAGE\xspace}
\newcommand{\myparagraph}[1]{\vspace{0.10cm}\noindent\hspace{0.1cm}\textbf{#1:}}

\newenvironment{promptquote}
  {\begin{quoting}[
      leftmargin=0.35em,
      rightmargin=0.35em,
      vskip=0.2\baselineskip
    ]\itshape}
  {\end{quoting}}

\begin{document}

\title{SAGE: Structured Agentic Graph Editing for Software Diagrams}

\settopmatter{authorsperrow=4}

\author{Tyler Sivertsen}
\affiliation{%
  \institution{Purdue University}
  \city{West Lafayette}
  \state{Indiana}
  \country{USA}
}
\email{tsiverts@purdue.edu}

\author{Neal Singh}
\affiliation{%
  \institution{Purdue University}
  \city{West Lafayette}
  \state{Indiana}
  \country{USA}
}
\email{sing1030@purdue.edu}

\author{James C. Davis}
\affiliation{%
  \institution{Purdue University}
  \city{West Lafayette}
  \state{Indiana}
  \country{USA}
}
\email{davisjam@purdue.edu}

\renewcommand{\shortauthors}{Sivertsen et al.}

\begin{abstract}
Software diagrams are difficult to edit through human-friendly interfaces because edits expressed in natural language must still preserve visual layout, editable structure, and semantic relationships.
As a step forward, we present \toolname{}, a browser-based tool for prompt-guided editing of Draw.io and Mermaid-style engineering diagrams. The tool maps diagrams into an editable graph representation, translates natural language requests into structured edit intents, analyzes those intents into graph-oriented operation steps, validates and repairs common Draw.io XML issues, and stores successful results as recoverable versioned artifacts. This design separates structured state management from model-driven interpretation, while acknowledging that some prompt-guided XML edits remain model-assisted. The tool also supports direct canvas editing and a secondary mask-based image-editing workflow. We evaluate the system using unit tests and a Kubernetes architecture case study, measuring structural validity, edit success, preservation of unrelated elements, and failure causes.

Demo video: \url{https://youtu.be/pB76i10GkMg}

Code: \url{https://github.com/Miller11k/Agentic-Figure-Drawing}
\end{abstract}


\maketitle

\section{Introduction}

Software evolves, and so should its engineering diagrams. Architecture diagrams, deployment diagrams, and workflow diagrams are common mechanisms for communicating system structure, behavior, and stakeholder-specific views~\cite{kruchten1995view,clements2010documenting,booch1999uml}. However, current drawing tools are either high-friction or under-controlled. Existing tools such as Draw.io and Mermaid support visual or textual diagram authoring~\cite{DrawioDocs,MermaidDocs}, but users must adapt their editing behavior to the tool's available affordances, such as dragging shapes, editing XML-like structure, or rewriting diagram text.
Recent systems leveraging language models can generate plausible visual outputs (\textit{e.g.},~\cite{OpenAIDocs,GeminiDocs}), but they treat diagrams as static images or unvalidated text.
Both approaches are insufficient for software engineering workflows where diagrams must remain editable, reusable, and structurally consistent. A useful diagram-editing tool must offer a human-friendly interface while preserving graph structure, maintaining editable elements, serializing back into reusable formats, and supporting recovery when an automatic edit is wrong.


Our tool, \toolname{}, addresses this problem through a hybrid architecture for structured, prompt-guided diagram editing. The tool maps Draw.io XML into an editable \textit{DiagramModel}, represents natural language requests as \textit{ParsedEditIntent} objects, analyzes those intents into target matches and graph-oriented operation steps, validates and repairs common serialized XML issues, and stores successful results as versioned artifacts.
Language-model reasoning is used for interpretation, target resolution, and some prompt-guided XML transformations.
Structured application logic performs graph edits, serialization, validation, repair, and recovery.

\myparagraph{Contributions}
  We describe the representations and algorithms that enable \toolname{} to combine human-oriented editing with structured diagram control and evaluate it with a case study. 
  More broadly, \toolname{} illustrates a pattern for agentic tools: use language models for interpretation, but constrain accepted changes through editable representations, deterministic logic, and validation guards.

\section{Tool Overview}

\begin{figure}[t]
    \centering
    \includegraphics[width=\linewidth]{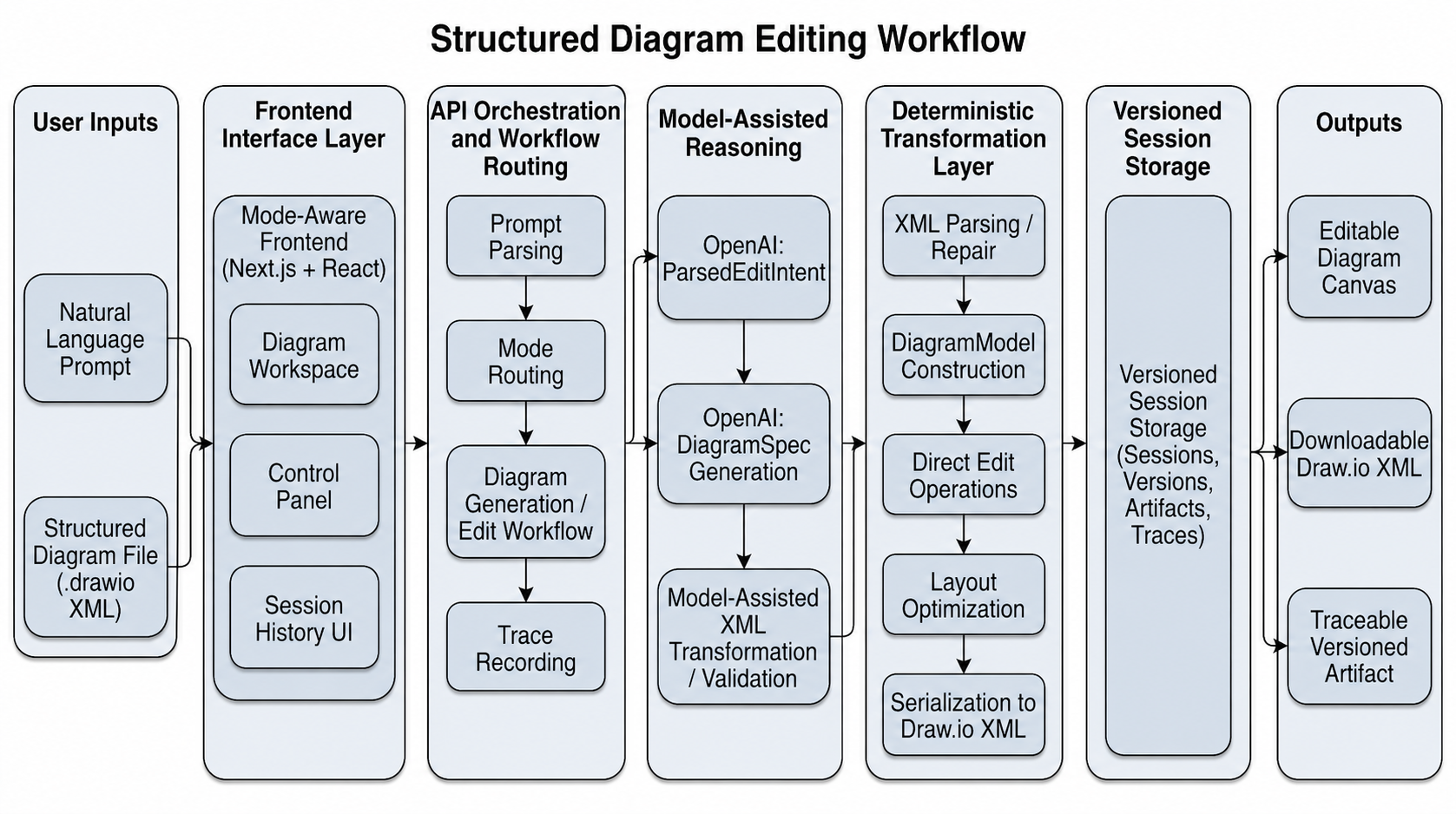}
    \caption{Diagram-editing workflow in \toolname{}.}
    \label{fig:structured-diagram-workflow}
\end{figure}

\toolname{} is a browser-based tool for creating, importing, editing, and exporting software engineering figures. The tool supports two primary use cases. In \textit{Diagram} mode, for example, users can import structured diagrams, generate diagrams from natural language, apply prompt-guided edits, and perform manual canvas edits. In \textit{Image} mode, users can upload or generate a legacy raster image and apply localized masked edits.
Figure~\ref{fig:structured-diagram-workflow} summarizes the structured diagram-editing workflow.

The central interface is a canvas backed by structured state. In diagram mode, users can select nodes and edges, move elements, edit labels, add or delete components, reconnect edges, inspect XML, and export the result. In image mode, users can draw masks for localized edits. Each operation creates a session step containing the prompt, workflow, intermediate representation, output artifact, and trace metadata. This makes edits non-destructive and inspectable. 

\section{Representations and Transformation Pipeline}

\toolname{} makes three design choices to make prompt-guided diagram editing compatible with structured software artifacts:
(1) diagrams are represented as editable graphs;
(2) natural-language requests are translated into explicit edit intents;
and
(3) generated results are validated before being accepted as recoverable artifacts.
Figure~\ref{fig:structured-diagram-workflow} shows the full \toolname{} workflow for diagram editing.
Together, these choices make the diagram an intermediate representation rather than a model output. The model can help interpret a user request, but the editable graph, operation steps, XML repair path, and versioned artifacts define what changes the system accepts.

\myparagraph{Editable graph model}
\toolname{} treats a software diagram as an editable graph rather than as a rendered image. The system normalizes imported Draw.io XML~\cite{DrawioDocs} into a \textit{DiagramModel}, which is the main internal representation used for editing, validation, rendering, and export. A \textit{DiagramModel} contains nodes, edges, labels, groups, geometry, style attributes, and source-target references. During import, the parser extracts diagram cells from Draw.io XML, separates vertex elements from connector elements, assigns stable identifiers, records geometry in a uniform coordinate system, and reconstructs graph relationships from connector metadata. Presentation details that do not affect editing are preserved when possible but are not used as the primary basis for transformation.

\myparagraph{Structured edit intents}
Natural language edits are represented as \textit{ParsedEditIntent} objects. An intent records the operation type, target selector, replacement content, insertion constraints, and optional layout constraints. For example, the instruction ``rename API Gateway to Edge Router'' becomes a rename operation with a label-based target selector and a replacement label. The instruction ``add a cache between the frontend and backend'' becomes an insert-between operation with two endpoint selectors and a new node specification. This representation is important because it separates prompt interpretation from diagram mutation. Model-driven reasoning is used to infer the user intent and resolve ambiguous references, but the actual diagram change is not applied directly from free-form model output.

\myparagraph{Execution planning}
The backend converts each \textit{ParsedEditIntent} into an editing analysis. This analysis binds the abstract intent to concrete diagram elements and records matched targets, selected operation steps, validation notes, and an execution route. The operation steps describe graph-oriented changes such as label updates, node insertion, edge creation, edge reconnection, styling, grouping, or layout adjustment. Preconditions are enforced at execution time by the direct-edit and workflow logic, for example by rejecting edits that reference missing nodes or invalid edge endpoints.

\myparagraph{Validation and repair}
After transformation, the system checks whether the result remains a valid editable diagram. The validator checks the basic shape of Draw.io-compatible XML, required root and layer cells, and connector references for edges that specify source and target endpoints. The repair logic handles common recoverable structural issues, such as empty outputs, missing Draw.io wrappers, and missing root or layer cells. Deeper semantic checks, including exhaustive duplicate identifier detection, label quality, finite geometry checks for all elements, and full group-membership validation, are only partially covered in the current prototype. If validation fails, the system prevents completed structured artifacts from being written as successful results. The current implementation creates a version step before validation.
Stronger transactional protection of the active version pointer remains future work.

\myparagraph{Separation of reasoning and execution}
This pipeline deliberately separates model-driven reasoning from structured execution and validation. The language model is used for tasks that require semantic interpretation, such as mapping ``database layer'' to a set of likely nodes or identifying which component the user meant by a partial name. Direct graph edits and diagram-spec serialization are deterministic after interpretation. Prompt-guided XML edits remain model-assisted, but their outputs are still passed through structured validation and repair before being accepted as usable artifacts. This gives the system a stronger failure model than a direct prompt-to-image workflow: even when the model participates in transformation, the result can be inspected, repaired for common structural issues, serialized, and versioned.

\myparagraph{Versioned recovery}
Each successful transformation creates a versioned session step. A step stores the original prompt, parsed intent, editing analysis, normalized diagram model, serialized XML artifact, rendered preview, and execution trace when those artifacts are available for the workflow. Reverting a diagram moves the current-version pointer to a previous step rather than rewriting history or deleting artifacts. This makes prompt-guided diagram editing inspectable and recoverable. It also supports evaluation because each run has explicit data for what the user requested, what the system inferred, what transformation was applied, and whether the resulting diagram remained structurally valid.

\section{Implementation}

\myparagraph{Implementation size}
\toolname{} is implemented as a full-stack web application using Next.js, React, and TypeScript~\cite{NextjsDocs,ReactDocs,TypeScriptDocs}. The implementation contains about 11,000 lines of source code excluding tests and fixtures, plus about 2,000 lines for the test and figure suite. The largest components are the frontend diagram editor and canvas UI, diagram representation, parsing and serialization, backend API routes, workflow orchestration, and model-provider integrations. Prisma is used to access a SQLite database for session metadata, version pointers, artifact references, workflow type, and trace summaries~\cite{PrismaDocs,SQLiteDocs}. Generated XML, rendered previews, masks, uploaded images, and model outputs are stored as local artifacts.

\myparagraph{Frontend architecture}
The frontend follows a model-view-\\controller pattern, corresponding to the interface layer shown in Figure~\ref{fig:structured-diagram-workflow}. In the diagram editor, the \textit{DiagramModel} is the model, the SVG canvas is the view, and user actions update structured state before rendering. This keeps the canvas from becoming the source of truth: the diagram model remains serializable, validatable, restorable, and exportable as Draw.io-compatible XML. The image workflow reuses the same session and artifact infrastructure, but operates over raster images and masks rather than graph objects.

\myparagraph{Backend architecture}
The backend is organized around workflow-specific API routes. Diagram import routes parse and normalize XML. Prompt-guided diagram routes call model-backed services to produce \textit{ParsedEditIntent} objects and resolve ambiguous targets. Transformation routes apply direct graph operations when possible, call model-assisted services for prompt-guided XML transformations, and run validation and repair before recording successful artifacts. Image routes package prompts, source images, and masks for provider-specific editing. Model providers, including OpenAI and optional Gemini workflows, are isolated behind service wrappers~\cite{OpenAIDocs,GeminiDocs}, so provider behavior is separated from storage, validation, artifact management, and session metadata.

\myparagraph{Development process}
LLM assistance was used during development, mainly for frontend iteration, automated test scaffolding, debugging, and localized bug fixes.
The authors designed the backend architecture, internal diagram representations, workflow structure, validation path, and artifact/versioning model.
All generated code was reviewed, revised, and integrated by the authors.

\section{Evaluation}

We evaluate \toolname{} by testing whether the system can reconstruct a realistic engineering diagram, convert it into an editable structured artifact, apply prompt-guided edits, and preserve diagram validity across revisions. The evaluation uses four metrics. \textit{Structural validity} checks whether the resulting artifact can be serialized as Draw.io-compatible XML and reloaded by the tool. \textit{Edit correctness} checks whether the requested change was applied to the intended target. \textit{Preservation} checks whether unrelated labels, nodes, and regions remain unchanged. \textit{Topology} checks whether the expected high-level connections remain present after editing.

A run is considered successful only if the output remains structurally valid and applies the intended edit without disrupting unrelated structure. This is stricter than checking visual plausibility. A diagram may look reasonable while still being unusable as an editable artifact if labels are missing, connectors are broken, identifiers are duplicated, or unrelated elements change during an edit.

\subsection{Diagram Reconstruction and Editing}

\begin{figure*}[t]
    \centering
    \begin{subfigure}[t]{0.32\textwidth}
        \centering
        \includegraphics[width=\linewidth]{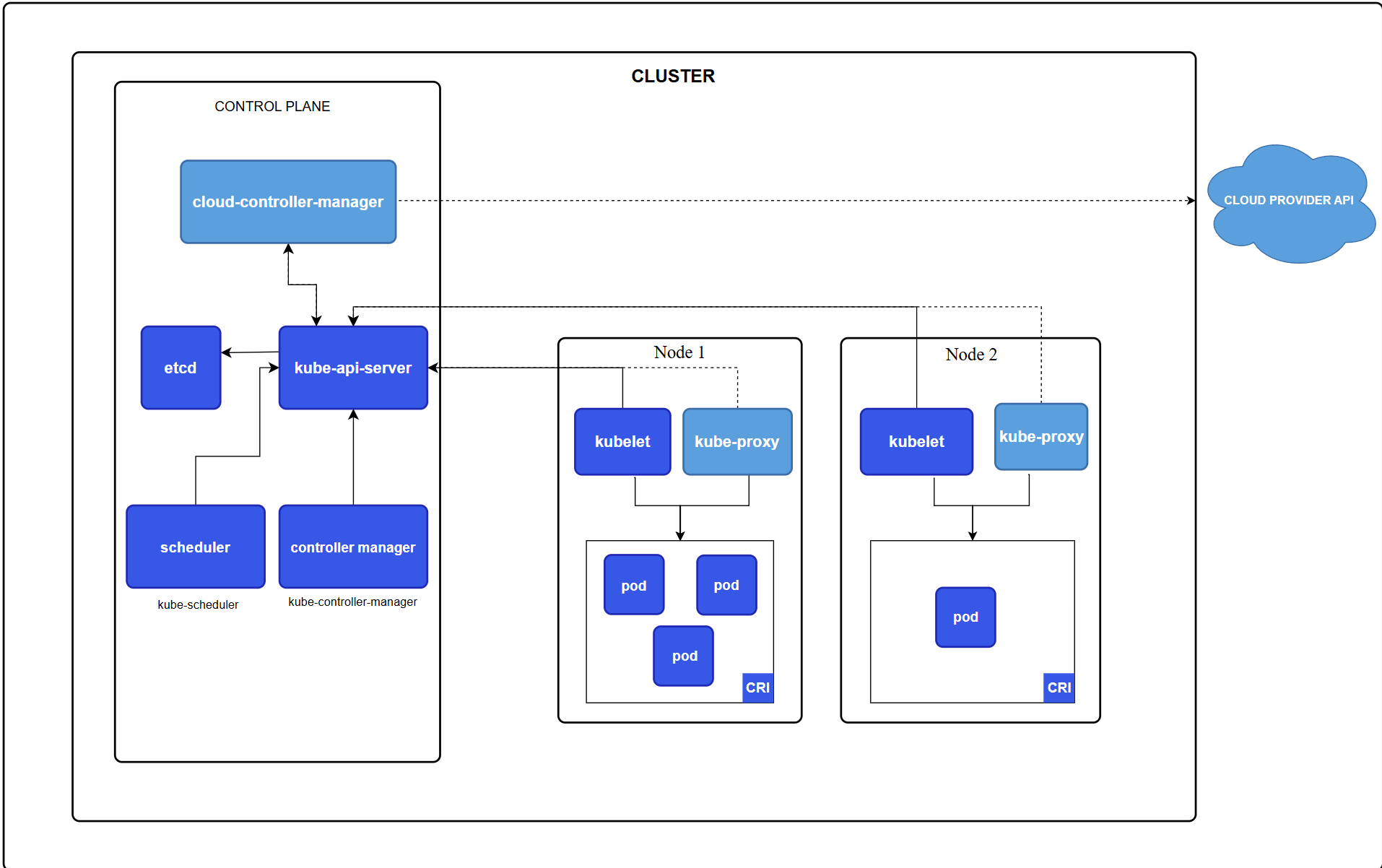}
        \caption{Reference input (from k8s docs)~\cite{KubernetesArchitecture}.}
        \label{fig:kubernetes-reference}
    \end{subfigure}
    \hfill
    \begin{subfigure}[t]{0.32\textwidth}
        \centering
        \includegraphics[width=\linewidth]{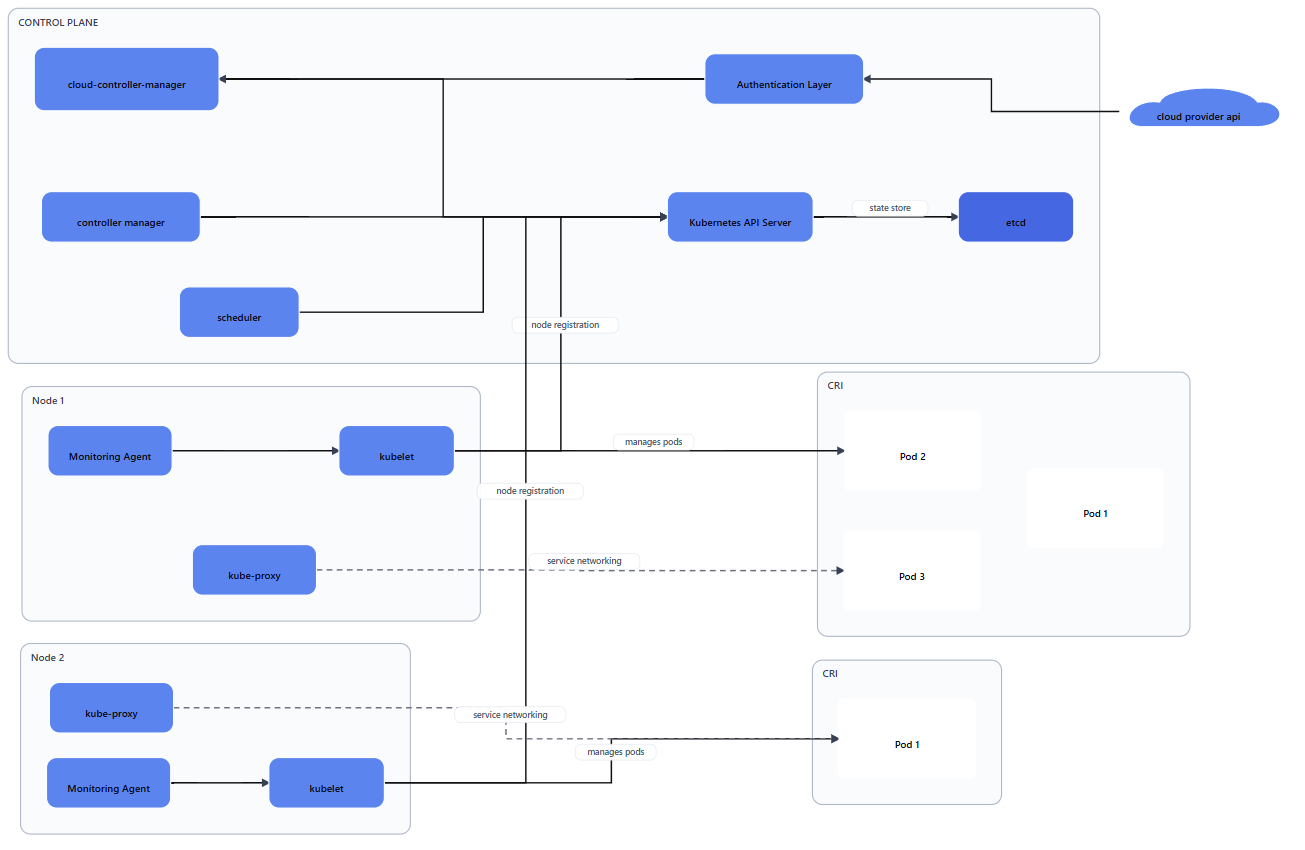}
        \caption{Structured edit result.}
        \label{fig:kubernetes-final-result}
    \end{subfigure}
    \hfill
    \begin{subfigure}[t]{0.32\textwidth}
        \centering
        \includegraphics[width=\linewidth]{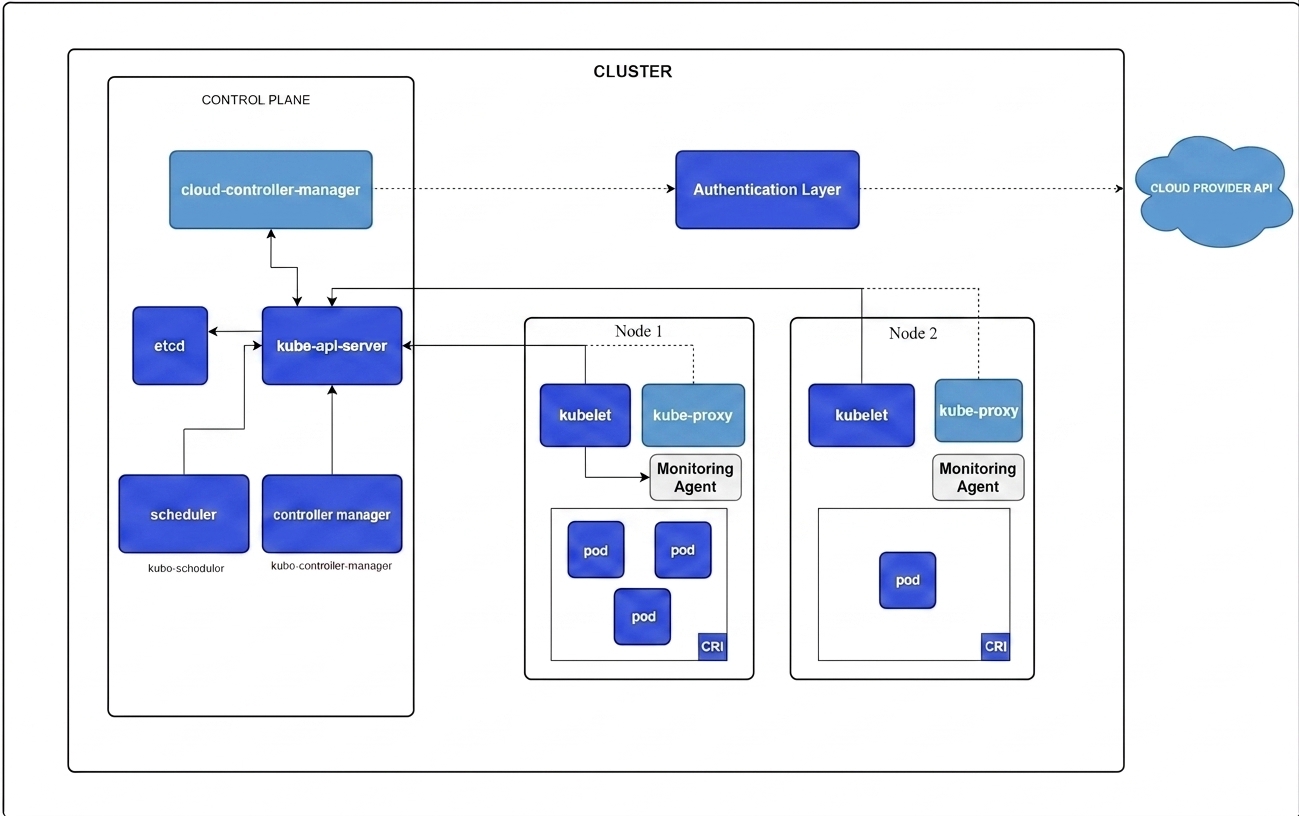}
        \caption{Image-editing result.}
        \label{fig:image-edit-kubernetes-final}
    \end{subfigure}
    \vspace{-0.25cm}
    \caption{Kubernetes case study artifacts: the reference diagram, final structured diagram-editing result, and final image-editing result. Larger versions of these three artifacts are included in the artifact repository.}
    \label{fig:kubernetes-case-study}
\end{figure*}

Our primary evaluation uses a Kubernetes cluster architecture diagram as a realistic reference input. The reference diagram is shown in Figure~\ref{fig:kubernetes-reference}. It contains a cluster boundary, a control plane, two worker nodes, labeled Kubernetes components, pod regions, and connector relationships. Because arbitrary Draw.io imports can depend on external icon libraries and project-specific styling conventions, this evaluation is a prompt-based reconstruction workflow rather than direct file import.

The system was given the following prompt:

\begin{promptquote}
\textit{Create a Kubernetes Cluster Architecture diagram. Within the cluster, there should be a CONTROL PLANE, Node~1, and Node~2. In the CONTROL PLANE there should be a cloud-controller-manager, etcd, kube-api-server, scheduler, and controller manager. Inside Node 1 there should be a kubelet, kube-proxy, and a CRI region with 3 pods. Inside Node 2 there should be a kubelet, kube-proxy, and CRI section with 1 pod. Connect the CLUSTER to a cloud provider API.}
\end{promptquote}

The reconstructed diagram was serialized as Draw.io-compatible XML, reloaded by the tool, and treated as the editable baseline for subsequent prompt-guided edits.
This demonstrates that \toolname{} can convert a reference figure into a structured \textit{DiagramModel} for subsequent inspection and modification.
Figure~\ref{fig:kubernetes-case-study} summarizes the reference input, final structured result, and final image-editing result; larger versions are provided in the artifact repository.

We then submitted three natural-language edit requests through the tool's prompt-edit interface. For each request, the system parsed the prompt, resolved the target diagram elements, generated the corresponding structured modification, and serialized the updated result as Draw.io-compatible XML:

\begin{itemize}
    \item Rename kube-api-server to Kubernetes API Server
    \item Add a Monitoring Agent inside each node and connect each Monitoring Agent to the kubelet in that node
    \item Insert a new component labeled Authentication Layer between cloud provider api and cloud-controller-manager. Route the connection so cloud provider API $\rightarrow$ Authentication Layer $\rightarrow$ cloud-controller-manager
\end{itemize}

These tasks exercise label mutation, node insertion, repeated target resolution, edge creation, and connector preservation.

In the Kubernetes case study, all structured edit outputs reloaded successfully. Preservation and topology were assessed by comparing each edited result against the reconstructed baseline.

\begin{table}[t]
\centering
\caption{Kubernetes edits from reconstructed baseline.}
\label{tab:kubernetes-results}
\begin{tabular}{lcccc}
\toprule
\textbf{Run} & \textbf{Correct} & \textbf{Pres.} & \textbf{Topo.} & \textbf{Failure} \\
\midrule
Reconstruction & Partial & --- & --- & Layout \\
Rename API server & Yes & Pass & Pass & --- \\
Add monitoring agents & Yes & Pass & Pass & --- \\
Insert auth layer & Yes & Pass & Pass & --- \\
\bottomrule
\end{tabular}
\end{table}

Table~\ref{tab:kubernetes-results} summarizes the results. The reconstruction was only partially correct because its layout differed from the reference, but it produced a reloadable structured artifact that could be used as the editable baseline. All three prompt-guided edits were applied to the intended targets, preserved unrelated diagram structure, and maintained the expected high-level topology.

The main limitation was layout quality. As prompts become more complex, diagrams can remain structurally valid while becoming visually crowded, producing overlapping labels or regions, or rendering slightly differently in the browser canvas than in Draw.io. These issues are usually correctable through minor manual edits, but they motivate stronger layout validation in future work.

\subsection{Secondary Image-Editing Evaluation}

We also evaluated the mask-based image-editing workflow as a secondary component of \toolname{}. Unlike the structured diagram workflow, this path operates directly on raster images and therefore does not expose an explicit graph representation, connector model, or validation scaffold. As a result, this evaluation focuses on whether semantically meaningful edits can be applied to an existing engineering figure, whether the major visual structure is preserved, and what limitations appear after repeated edits.

We used the Kubernetes architecture image from the diagram benchmark as the image-editing input. Starting from the original image, we applied two sequential semantic edits: inserting an \texttt{Authentication Layer} between the \texttt{cloud-controller-manager} and the \texttt{CLOUD PROVIDER API}, and adding a \texttt{Monitoring Agent} inside both Node~1 and Node~2. The result is shown in Figure~\ref{fig:image-edit-kubernetes-final}.

These results show that the image-editing workflow can support nontrivial semantic diagram modifications. Both requested edits were applied successfully, and the resulting image remained understandable as a Kubernetes architecture diagram. However, repeated edits introduced visible degradation in some areas, including reduced visual consistency, slight layout drift, and typography changes. In contrast to the structured diagram-editing pipeline, the image-editing workflow does not use an explicit intermediate representation or deterministic validation checks. Consequently, semantic correctness, connector preservation, and layout stability are left to the image model itself.

This secondary evaluation complements the structured diagram benchmark by illustrating a tradeoff between the two workflows. The structured diagram pipeline provides stronger guarantees through explicit representations, deterministic transformations, and validation, while the image-editing pipeline offers more flexible visual editing at the cost of weaker correctness guarantees and increased susceptibility to cumulative degradation across repeated edits.

\vspace{-0.25cm}
\section{Discussion}
\label{sec:discussion}
\toolname{} illustrates two broader implications for software engineering. First, prior work on architecture documentation emphasizes that architecture descriptions must communicate system structure in a form that stakeholders can understand, use, and maintain over time~\cite{clements2010documenting,kruchten1995view}. In practice, documentation, including system models and diagrams, often becomes stale as the implementation changes. Our evaluation showed that \toolname{} supports targeted edits such as renaming components, inserting services, reconnecting dependencies, or updating deployment structure, which can lower the cost of keeping diagrams current.



Second, and more forward-looking, structured diagram editing can make human--agent collaboration more inspectable than text-only interaction.
Many agentic software engineering workflows are mediated through prompts, code diffs, logs, and natural-language explanations.
These interfaces are useful, but are poorly matched to architectural reasoning, where engineers think in terms of components, boundaries, flows, ownership, and invariants.
Automatically editable diagrams offer a distinctive \textit{intermediate representation} to facilitate human-AI teaming. 

These implications also clarify the limits of the current prototype.
\toolname{} is strongest when the desired edit can be represented as a graph-oriented transformation over nodes, edges, labels, groups, and layout constraints.
It is weaker when correctness depends on domain-specific architectural semantics, aesthetic layout quality, or information that is not present in the diagram.

\section{Conclusion}

\toolname{} demonstrates that prompt-guided diagram editing is feasible for software engineering when diagrams are treated as structured, editable artifacts rather than static images, and when agentic generation is constrained by validation, repair, versioning, and recoverable state.
Our formative evaluation shows that the prototype can reconstruct and revise a realistic architecture diagram, apply targeted edits, preserve unrelated structure, and export reloadable Draw.io-compatible artifacts, while also exposing limitations in layout quality, semantic validation, and benchmark breadth.
More broadly, diagrams need not remain passive documentation that drifts behind the system: with appropriate structure and controls, they can become maintainable artifacts through which engineers review, guide, and constrain agentic architectural changes.


\bibliographystyle{ACM-Reference-Format}
\bibliography{agentic_diagram_acm_references}

@misc{NextjsDocs,
  author       = {{Vercel}},
  title        = {Next.js Documentation},
  year         = {2026},
  howpublished = {\url{https://nextjs.org/docs}},
}

@misc{ReactDocs,
  author       = {{Meta Open Source}},
  title        = {React Documentation},
  year         = {2026},
  howpublished = {\url{https://react.dev/}},
}

@misc{TypeScriptDocs,
  author       = {{Microsoft}},
  title        = {TypeScript Documentation},
  year         = {2026},
  howpublished = {\url{https://www.typescriptlang.org}},
}

@misc{PrismaDocs,
  author       = {{Prisma Data, Inc.}},
  title        = {Prisma Documentation},
  year         = {2026},
  howpublished = {\url{https://www.prisma.io/docs}},
}

@misc{SQLiteDocs,
  author       = {{SQLite Consortium}},
  title        = {SQLite Documentation},
  year         = {2026},
  howpublished = {\url{https://www.sqlite.org}},
}

@misc{OpenAIDocs,
  author       = {{OpenAI}},
  title        = {OpenAI API Reference},
  year         = {2026},
  howpublished = {\url{https://developers.openai.com}},
}

@misc{GeminiDocs,
  author       = {{Google AI for Developers}},
  title        = {Gemini API Docs},
  year         = {2026},
  howpublished = {\url{https://ai.google.dev}},
}

@misc{DrawioDocs,
  author       = {{JGraph Ltd.}},
  title        = {draw.io Documentation},
  year         = {2026},
  howpublished = {\url{https://www.drawio.com/doc/}},
}

@misc{MermaidDocs,
  author       = {{Mermaid}},
  title        = {Mermaid Documentation},
  year         = {2026},
  howpublished = {\url{https://mermaid.js.org/intro/}},
}

@misc{KubernetesArchitecture,
  author       = {{The Kubernetes Authors}},
  title        = {Kubernetes Components},
  year         = {2026},
  howpublished = {\url{https://kubernetes.io/docs/concepts/architecture/}},
}

@article{kruchten1995view,
  author = {Kruchten, Philippe},
  title = {The 4+1 View Model of Architecture},
  journal = {IEEE Software},
  volume = {12},
  number = {6},
  pages = {42--50},
  year = {1995},
  doi = {10.1109/52.469759}
}

@book{clements2010documenting,
  author = {Clements, Paul and Bachmann, Felix and Bass, Len and Garlan, David and Ivers, James and Little, Reed and Merson, Paulo and Nord, Robert and Stafford, Judith},
  title = {Documenting Software Architectures: Views and Beyond},
  edition = {2},
  publisher = {Addison-Wesley Professional},
  year = {2010},
  isbn = {9780321552686}
}

@book{booch1999uml,
  author = {Booch, Grady and Rumbaugh, James and Jacobson, Ivar},
  title = {The Unified Modeling Language User Guide},
  publisher = {Addison-Wesley},
  year = {1999},
  isbn = {0201571684}
}

\end{document}